\documentstyle[aps,preprint]{revtex}  
\begin{document}
%\draft
\tighten
\preprint{\vbox{Submitted to Physics Letters B 
                \hfill YUM 96-14\\
\null \hfill  SNUTP 96-067 }}
\title{QCD Sum Rule for  S$_{11}$(1535)}
\author{Su Houng Lee$^{a,b}$\footnote{E-mail:suhoung@phya.yonsei.ac.kr} and
Hungchong Kim$^{a}$\footnote{E-mail:hung@phya.yonsei.ac.kr}}
\address{$^a$ Department of Physics, Yonsei University, Seoul, 120-749, Korea
\\
 $^b$ GSI, Postfach 11 05 52, D-64220 Darmstadt, Germany}
\date{\today}
\maketitle
\begin{abstract}
We propose a new interpolating field for S$_{11}$(1535) to determine
its mass from QCD sum rules.  In the nonrelativistic limit, this 
interpolating
field dominantly reduces to two quarks in the s-wave state and one quark in 
the p-wave state.  An optimization procedure, which makes use of a duality
relation,  yields the interpolating field
which overlaps strongly with the negative-parity baryon and at the same
time does not couple at all to the low lying positive-parity baryon.  
Using this interpolating field and applying the conventional QCD sum rule
analysis, we find that the mass of S$_{11}$ is reasonably close to the 
experimentally known  value, even though the precise determination depends
on the poorly known quark-gluon condensate.  Hence our interpolating field 
can be used to investigate the spectral properties of S$_{11}$(1535). 

\vspace{0.8cm}


\end{abstract}  


The negative parity nucleon resonance S$_{11}$(1535) has always
been an object of study when  constructing models of the baryon.  
A good fit to its mass and decay widths provides further test of the
validity of the model, which in return provides a deeper 
understanding of the structure of the resonance itself.

Recently, in relation to the present and planned experiments of 
$\eta$-production on a nucleon from electromagnetic
or hadronic probes at MAMI (Mainz)\cite{mami}, ELSA (Bonn)\cite{elsa}
and CEBAF, 
there has been a renewed interest in S$_{11}$.  Because
the mass of S$_{11}$ is just above the $\eta$N threshold,
$\eta$-productions in these experiments are dominated by s-channel 
S$_{11}$ resonance contribution.   As a consequence, a 
detailed study of its properties, such as  
the $\eta$-nucleon-S$_{11}$ coupling and transition form factors,  
can provide important constraints in the analysis of these experimental data.
So far, most theoretical works\cite{eta_the}  have been focused on 
reproducing
the experimental data by using the empirical parameters related to the 
S$_{11}$, which have been explicitly introduced in their
effective models.  To make a connection between these effective models with 
presumably the fundamental theory of QCD,  it is important to understand how
the spectral properties of S$_{11}$ are generated
from  QCD order parameters. In this way, 
phenomenological aspects of S$_{11}$ can be predicted
rather model-independently.   In this letter, as a starting point for this
kind of analysis, we will introduce a new interpolating field 
for the S$_{11}$ suitable for conventional QCD sum rule analysis.  

The QCD sum rule\cite{SVZ} is a useful tool to study the spectral properties 
of the hadrons.  It has been widely used in a number of occasions; 
in calculating properties of hadrons, coupling  strength between hadrons, 
form factors\cite{qsr} and recently medium dependent 
properties\cite{Tom1,Lee1}.  The starting point of 
the QCD sum rule is to introduce an appropriate interpolating field for the
hadron of concern.  In general,  
the interpolating field is constructed from quark and gluon  
fields, whose specific form is  usually determined by considering the 
hadron's quantum numbers and its nonrelativistic quark wave functions. 
Then the two-point correlation function 
of such interpolating field is introduced and
calculated in the operator product expansion.  By matching the resulting
correlation function with 
its phenomenological counterpart, one can make prediction on
the hadron's spectral properties in terms of the QCD order parameters.   
Clearly, the success of the prediction is highly dependent upon the choice of
the interpolating field.
 
Recently, Jido Kodama and Oka\cite{jido} have proposed
a technique to separate out the contribution of the negative-parity baryon
from QCD sum rules 
using the usual interpolating fields of the nucleon.  They succeeded in
obtaining a reasonable value  for the  mass of S$_{11}$ by adjusting
the QCD input parameters and at the same time optimizing a linear 
combination of the two independent nucleon interpolating fields. 
However, there one has to take the difference between the two independent sum 
rules of
the nucleon each proportional to the Dirac structure of $1$ and $\not\!q$
so as to
subtract out the nucleon contribution, which has a large overlap with the 
current.   
This fact makes it difficult to investigate other properties, 
such as the form factors or the couplings,  of the S$_{11}$ 
resonance within the conventional sum rule approach.  
The problem is that the two independent nucleon currents 
used so far either couple strongly or weakly to both the nucleon and 
the low-lying resonances at the same time,  
such that it is difficult to select out only the negative
parity part of the resonances in either case.  To overcome this difficulty,
we will construct a current that couples weakly to the nucleon and strongly
to the S$_{11}$. 
It should be noted that the reason why the nucleon couples 
strongly with the commonly used {\it Ioffe current} is that, this current 
has a strong
overlap with the nonrelativistic quark wave function of the nucleons in which 
all quarks are in the s-wave state.   
Therefore, we will  construct a current whose 
nonrelativistic limit has a strong overlap with two quarks in the s-wave 
state and one quark in the p-wave state.   This is the picture for S$_{11}$ 
both in the  
bag model or nonrelativistic quark model.  This implies that we have to go
beyond the two independent nucleon sum rule and introduce an appropriate 
covariant derivative in the current, for which there are many choices.  
We will choose a minimal approach and construct our current so that it does
not couple to the low lying positive-parity nucleon but couples strongly to 
the negative-parity baryon. 
As will be shown by QCD sum rule analysis, our current gives a reasonably 
stable plateau for its mass 
in the Borel curve.  
Using this current, it is possible to apply conventional sum rule approaches
to investigate other properties of the  S$_{11}$ resonance, which will be 
a subject of future study~\cite{hung}
                                                                     
First, we start with a short discussion of the proton
interpolating field. The interpolating field of the proton is 
constructed from two u-quarks and one d-quark by assuming that
all three quarks are in the s-wave state. Specifically,
one up and the down quarks are combined into an isoscalar diquark and
the other up quark is attached to the diquark so that the quantum numbers of 
the proton are carried by the attached up quark.  The proton should be a
color singlet and its interpolating field should  transform as
the spinor under the parity transformation.   Then, it is found that 
there are two interpolating fields possible~\cite{ioffe,Esp,griegel},
\begin{eqnarray}
\eta_1 = \epsilon_{abc} (u_a^{\rm T} C d_b) \gamma_5 u_c\ ,\\
\eta_2 = \epsilon_{abc} (u_a^{\rm T} C \gamma_5 d_b) u_c\ ,
\end{eqnarray}
where $\eta_2$ contains the valence quark wave function in the
nonrelativistic limit.   
In general, the interpolating field for a proton is an arbitrary 
linear combination of these fields, 
\begin{eqnarray}
\eta(t)=2 \epsilon_{abc}[\ t\ (u_a^{\rm T} C d_b) \gamma_5 u_c
+(u_a^{\rm T} C \gamma_5 d_b) u_c]\ .
\label{nucl1}
\end{eqnarray}
For $t$ equal to $-1$, 
the resulting interpolating field can be shown to have no  direct 
instanton contribution and, at the same time  
couple strongly  to the chiral symmetry breaking effects~\cite{ioffe}.   

S$_{11}$ is the lowest resonance of the nucleon  with negative 
parity.  As suggested by the bag model~\cite{chodos}
or nonrelativistic quark model~\cite{isgur}, one quark in S$_{11}$
is believed to be in the p-wave state relative to the other two quarks.
We want to construct a current, which in the nonrelativistic limit
overlaps largely with this quark field configuration.  One way to
accomplish this is to put a covariant derivative $z \cdot D$
to one of the quarks in  Eq.~(\ref{nucl1}).  Here, we have introduced a 
four vector $z^\mu$ such that  $z \cdot q=0$ and $z^2=-1$.  
The property of $z^\mu$ is chosen to make the  covariant 
derivative orthogonal to the four vector
$q$ carried by the S$_{11}$ so that in the rest frame, $z \cdot D$
reduces to the derivative in the space direction.
Of course, there are
three possible choices for putting in the derivative inside the nucleon
current.  
In this work, we take the interpolating field
with the covariant derivative acting on the d-quark instead of other two 
u-quarks :  
\begin{eqnarray}
\eta_{N^-} (t)=2 \epsilon_{abc}[\ t\ (u_a^{\rm T} C (z \cdot D)d_b) 
\gamma_5 u_c
+(u_a^{\rm T} C \gamma_5 (z \cdot D) d_b) u_c]\ .
\label{nstar}
\end{eqnarray}
This choice is preferred because
the resulting correlator gets nonzero contribution from
the lowest-order chiral breaking term, $\langle {\bar q} q \rangle$, which
is  important in the nucleon sum rule.
  
With this current, we now  
move on to calculate the time-ordered correlation function defined as 
\begin{eqnarray}
\Pi(q) = \int dx^4 e^{iq\cdot x} i\ \langle 0 | {\rm T} [ \eta_{N^-} (x) 
{\bar \eta_{N^-}} (0)] | 
0 \rangle\ , 
\label{cor}
\end{eqnarray}
where $|0\rangle$ denotes the QCD vacuum.
Implementing the properties of $z_\mu$, it is easy to show that the
correlation function should have the following form,
\begin{equation}
\Pi(q) = \Pi_1(q^2, z^2) + \Pi_q(q^2,z^2) \not\!q \ .
\label{pi}
\end{equation}
Calculation of these two scalar functions is carried out readily 
in the operator product expansion.  
After some manipulations, we obtain up to dimension 8 operators, 
\begin{eqnarray}
\Pi^{\rm ope}_q (q) &=&  -{ q^6 {\rm ln} (-q^2) \over 2^{10}\times 3^2 \times
                         5\times \pi^4 } (21 t^2 + 10 t + 21)
                        - { q^2 {\rm ln} (-q^2) \over 2^{11}\times 
                         9 \times \pi^2 } (45 t^2 + 10 t + 45)
                        \langle \frac{\alpha_s}{\pi} {\cal G}^2
                        \rangle \nonumber \\
                    &&- { 1 \over 12 q^2} \langle {\bar q} q \rangle
                        \langle g_s {\bar q} \sigma \cdot {\cal G} q \rangle
                        (t^2 + t)\label{ope1}\ ,\\
\Pi^{\rm ope}_{1} (q) &=& {q^4 {\rm ln} (-q^2) \over 2^5 \times \pi^2 }(t^2-1)
                          \langle {\bar q} q \rangle -
                     { q^2 {\rm ln} (-q^2)
                     \over 2^7  \times \pi^2}
                     (3 t^2 + 2 t -5)
                    \langle g_s {\bar q} \sigma \cdot {\cal G} q \rangle
\nonumber \\
                    && + { {\rm ln} (-q^2) \over 3 \times 2^6}
                         (t^2-1)
                         \langle \frac{\alpha_s}{\pi} {\cal G}^2 \rangle
                         \langle {\bar q} q \rangle\ .
\label{ope2}
\end{eqnarray}
Note that, in calculating these, we
have made use of the properties, $z\cdot q =0$ and $z^2 =-1$.
Few remarks are in order. 
The Wilson coefficients of 
the dimension six condensate ($\langle {\bar q}q \rangle^2$) in
$\Pi^{\rm ope}_q (q)$  can be shown to 
be zero.  We added the contribution from the dimension eight
operator  ($\langle {\bar q}q \rangle
\langle g_s {\bar q} \sigma \cdot {\cal G} q \rangle$), as they contribute
in tree graph and hence expected to be the important power correction
for $\Pi^{\rm ope}_{q}$.
For $\Pi_1(q^2,z^2)$, the leading contribution comes from dimension three
condensate, $\langle {\bar q}q \rangle$, which constitutes the important
chiral symmetry breaking operator in predicting the nucleon mass in 
the nucleon sum rule. 
In addition, another chiral breaking operator,
$\langle g_s {\bar q} \sigma \cdot {\cal G} q \rangle$, contributes to
the S$_{11}$ sum rule. This operator does not enter to the usual
nucleon sum rule where {\it Ioffe} current is employed. 
However, as will be discussed below, our prediction for
S$_{11}$ mass is sensitive to the value of this dimension-five condensate.
Note also that the factorization hypothesis has been employed in the 
calculation of higher dimensional operators.
We did not include the three-gluon condensate 
$\langle {\cal G}^3 \rangle$ which is an order of $\alpha_s^{3/2}$.

After Borel transformation, Eqs.~(\ref{ope1})and~(\ref{ope2}) become,
\begin{eqnarray}
{\hat \Pi}^{\rm ope}_q (M)&=& {  M^8 \over 2^9\times 3 
                              \times 5 \times \pi^4}(21 t^2+10t+21) 
                              + { M^4 \over 2^{11}\times 9 \times 
                             \pi^2}(45 t^2+10t+ 45) 
           \langle {\alpha_s \over \pi} {\cal G}^2 \rangle \nonumber \\
                        &&+ {1 \over 12} (t^2+t) \langle {\bar q} q \rangle
                          \langle g_s {\bar q} \sigma \cdot {\cal G} q \rangle
                          \label{bope1}\ , \\
{\hat \Pi}^{\rm ope}_{1} (M) &=&  -{M^6 \over 2^4 \pi^2}
               \langle {\bar q} q \rangle (t^2-1)+{ M^4 \over 2^7 \pi^2}
                                 \langle g_s {\bar q} \sigma \cdot {\cal G} 
                                  q \rangle\ (3 t^2 + 2 t -5)\nonumber \\
                            &&- { M^2 \over 192}
                         \langle \frac{\alpha_s}{\pi} {\cal G}^2 \rangle
                         \langle {\bar q} q \rangle (t^2-1)\ ,
\label{bope2}
\end{eqnarray}
where $M$ denotes the Borel mass.

%%%%%%%%%%%%%%%%%%%%%%%%%%%%%%%%%%%%%%%%%%%%%%%%%%%%%%%%%%%%%%%%%%%%%%%

For the phenomenological side of the sum rules, first we consider 
the matrix element of the interpolating field [Eq.~(\ref{nstar})]
between the QCD vacuum and a baryon state with momentum $q$ and spin $s$.
Due to the specific form of the field, we should have
\begin{equation}
\langle 0 | \eta(0) | q,s \rangle = 
\not \!z \lambda_B\ u(q,s)\ ,
\label{mat}
\end{equation} 
where $u(q,s)$ is a Dirac spinor and 
$\lambda_B$ is the strength with which the baryon couples to
the interpolating field.  Normally, the interpolating field couples
not only to the positive-parity baryon but also to the negative-parity
baryon.  This implies that, when combined with Eq.~(\ref{mat}), 
the phenomenological form of the correlator should be~\cite{jido} 
\begin{eqnarray}
\Pi^{\rm phen} (q) &=&  - \Bigr [ \lambda_N^2 {  \not\!q - M_N \over
                  q^2 -M_N^2 +i\epsilon } 
                  + \lambda_{N^-}^2 {\not\!q + M_{N^-} \over
                 q^2 -M_{N^-}^2 + i\epsilon } \Bigl ]+~{\rm continuum}\ .
\label{phe}
\end{eqnarray}
Here we retain  the lowest resonance of positive parity which we
identify as the nucleon($N$),  and
the lowest resonance of negative parity which we identify 
as S$_{11}$ ($N^-$).  We have put all other resonances with higher masses 
into two separate continua, one for the positive-parity resonances which we 
denote by $s_+$, the other for the negative-parity resonances which we denote
by $s_-$. 
$\lambda_N$ and $\lambda_{N^-}$  represent the strengths with which 
our interpolating field
couples to nucleon and S$_{11}$, respectively while the corresponding
masses are denoted by $M_N$ and $M_{N^-}$.

In general, the couplings, $\lambda_N$ and $\lambda_{N^-}$
are functions of $t$.  To obtain the sum rule for S$_{11}$, 
the parameter $t$ need to be chosen such a way that
the resulting interpolating field does not couple to the
positive-parity baryons while strongly couples to S$_{11}$.   
This can be achieved by applying the finite energy sum rule as described
in Ref.~\cite{jido}.  We will come back to this procedure later.  
However, once  $t$ is chosen this way, there is no contribution from
the nucleon, and therefore one can derive the sum rule 
for S$_{11}$ in the conventional way.  
Following the usual steps to include the continuum [see for 
example Ref.~\cite{griegel}] and taking the Borel transformation of the
resulting expressions,   
we get the phenomenological side of the sum rule assuming two different 
continuum thresholds,
\begin{eqnarray}
{\hat \Pi}_q^{\rm phen} (M) &=& \lambda^2_{N^-} e^{-M^2_{N^-}/M^2}
                          +{ (21t^2+10t+21) \over 2^9 \times 3
                          \times 5 \times \pi^4} \frac{1}{2}
(M_{s_-}^8+M_{s_+}^8)
\nonumber \\
&& + { \langle {\alpha_s \over \pi} {\cal G}^2 \rangle
                        \over 2^{11}\times 9\times \pi^2}(45 t^2+10t+45)
                         \frac{1}{2}(M_{s_-}^4+M_{s_+}^4) \ ,
\nonumber \\  
 {\hat \Pi}_1^{\rm phen} (M) &=& \lambda^2_{N^-} M_{N^-} e^{-M^2_{N^-}/M^2}
- { \langle {\bar q} q \rangle \over 2^4 \pi^2} (t^2-1)
 \frac{1}{2}(M_{s_-}^6+M_{s_+}^6)
\nonumber \\
& & -{ \langle g_s {\bar q} \sigma \cdot {\cal G} q
                       \rangle \over 2^7 \pi^2} 
                            (3 t^2+2t - 5)
                         \frac{1}{2}(M_{s_-}^4+M_{s_+}^4)
\nonumber \\
 && + { (t^2 -1) \over 192}
\langle {\alpha_s \over \pi} {\cal G}^2 \rangle \langle {\bar q} q \rangle
\frac{1}{2}(M_{s_-}^2+M_{s_-}^2) \ ,
\end{eqnarray}
where we have defined,
\begin{eqnarray}
M^2_{s_{\pm}} &=& M^2  e^{-s_{\pm}/M^2}  \ ,\\
M^4_{s_{\pm}} &=& M^4 (1 + {s_{\pm} \over M^2})e^{-s_{\pm} /M^2}  \ ,\\
M^6_{s_{\pm}} &=& M^6 (1 + {s_{\pm} \over M^2} + 
{s_{\pm}^2 \over 2 M^4})e^{-s_{\pm} /M^2}  \ ,\\
M^8_{s_{\pm}} &=& M^8 (1 + 
{s_{\pm} \over M^2}+{s_{\pm}^2 \over 2 M^4}+{s_{\pm}^3 \over 6 M^6})
e^{-s_{\pm}/M^2}\ .
\end{eqnarray}

By equating these phenomenological side with the OPE side of the sum rule
[Eqs.~(\ref{bope1}), (\ref{bope2})] and taking the ratio,
we obtain the sum rule for $M_{N^-}$,
\begin{eqnarray}
M_{N^-} &=& \Big[ -{\Delta(M_{s_-}^6)+ \Delta(M_{s_+}^6) \over 2}
(t^2-1)\langle 
  {\bar q} q \rangle+ 
{\Delta(M_{s_-}^4)+ \Delta(M_{s_+}^4) \over 16}(3t^2+2t-5)\langle 
 g_s {\bar q} \sigma \cdot {\cal G}
                         q \rangle\  \nonumber\ \\
     &&
            -{\Delta(M_{s_-}^2)+ \Delta(M_{s_+}^2) \over 24} \pi^2 (t^2-1) 
                 \langle {\alpha_s \over \pi} {\cal G}^2 \rangle 
                 \langle {\bar q} q \rangle 
\Big]  \nonumber \\
&\times& \Big[ { \Delta(M_{s_-}^8)+ \Delta(M_{s_+}^8) \over 960 \pi^2} 
  (21 t^2 +10t +21) \nonumber\ \\ 
      && +{\Delta(M_{s_-}^4)+ \Delta(M_{s_+}^4) \over 256 \times 9}\langle 
  {\alpha_s \over \pi} {\cal G}^2 \rangle
       (45 t^2 + 10 t + 45)  \nonumber\ \\
    && + {4\pi^2\over 3} (t^2+t) \langle {\bar q} q \rangle 
                          \langle g_s {\bar q} \sigma \cdot {\cal G}
                          q \rangle\  \Big]^{-1},
\label{mass}
\end{eqnarray}
where we have defined
\begin{eqnarray}
\Delta(M^2_{s_{\pm}}) &=& M^2 -M_{s_{\pm}}^2 \ ,\\
\Delta(M^4_{s_{\pm}}) &=& M^4 -M_{s_{\pm}}^4 \ ,\\
\Delta(M^6_{s_{\pm}}) &=& M^6 -M_{s_{\pm}}^6 \ ,\\
\Delta(M^8_{s_{\pm}}) &=& M^8 -M_{s_{\pm}}^8 \ .
\end{eqnarray}

Now, the QCD parameters appearing in Eq.~(\ref{mass}) should be 
properly chosen to predict $M_{N^-}$.  The quark condensate can be deduced 
 from Gell-Mann$-$Oakes$-$Renner relation combined
with the current quark masses estimated from chiral perturbation
theory~\cite{griegel}:
\begin{equation}
\langle {\bar q} q \rangle = - (230 \pm 20\ {\rm MeV})^3\ .
\label{qq}
\end{equation}
The gluon condensate can be estimated either from leptonic decays of vector 
mesons or from the charmonium spectrum~\cite{SVZ}:
\begin{equation}
\langle{\alpha_s \over \pi} {\cal G}^2 \rangle = (350 \pm 20\ {\rm MeV})^4 \ .
\label{gluon}
\end{equation}

The most important parameter for our prediction is the quark-gluon
condensate which is usually expressed in terms of the quark condensate
$\langle {\bar q} q \rangle$: 
\begin{equation}
\langle g_s {\bar q} \sigma \cdot {\cal G} q \rangle\ \equiv  
2\lambda^2_q \langle {\bar q} q \rangle\ . 
\label{qgc}
\end{equation}
Here, the parameter $\lambda_q^2$, which represents the
average virtual momentum of vacuum quarks, is not precisely known. It 
has been estimated from a number of studies; 
$\lambda^2_q = 0.4 \pm 0.1$ GeV$^2$ from the standard QCD 
sum rule estimate~\cite{Ovc}, $\lambda^2_q= 0.7$ GeV$^2$  
from QCD sum-rule analysis of the pion form factor~\cite{bak}, 
$\lambda^2_q= 0.55$ GeV$^2$ from lattice calculations~\cite{kremer}, 
and somewhat larger value from the instanton liquid model.  
In our discussion, we will investigate the sensitivity of our prediction
by changing  $\lambda^2_q$ between $0.4 - 1$ GeV$^2$.

Given the QCD parameters,
we need to choose an optimal $t$ such that 
the interpolating field does not couple to the positive-parity
nucleon at all while maximally couples to the negative-parity nucleon.
To do this, we use the sum rule 
derived by JKO in Ref.~\cite{jido}, where they have used the ``old-fashioned"
correlation function defined by
\begin{eqnarray}
\Pi(q)= i \int d^4x e^{iqx} \theta(x_0) \langle 0| \eta_{N^-} (x) 
\eta_{N^-} (0) |0 \rangle\ .
\end{eqnarray}
This function is analytic in the upper-half region of the complex $q_0$ plane.
Therefore, using the nucleon current and assuming,
\begin{eqnarray}
{\rm Im}\ \Pi(q_0, {\bf q}=0)=\gamma_0 A(q_0)+B(q_0),
\end{eqnarray}
one can derive the following sum rules for a large energy $Q$,
\begin{eqnarray}
\int_0^{Q} dq_0 [A^{\rm ope}(q_0)-A^{\rm phen}(q_0)] W(q_0)
&=& 0 \nonumber \\
\int_0^{Q} dq_0 [B^{\rm ope}(q_0)-B^{\rm phen}(q_0)] W(q_0)
&=& 0\ ,
\label{jkos}
\end{eqnarray}
where $W(q_0)$ is the weighting function.  The advantage of using the
``old-fashioned'' sum rule is that it allows us to
determine the optimal $t$ independent of 
baryon masses which we want to calculate ultimately.

As stated above [Eq.(\ref{phe})], 
we assume that our phenomenological side has explicit 
contributions from the nucleon ($N$) and the $S_{11} (N^-)$.
That is, we have
\begin{eqnarray}
{1 \over \pi} {\rm Im}\ [\Pi^{\rm phen} (q_0)] &=& 
\lambda_N^2 \frac{\gamma_0-1}{2} 
\delta(q_0-M_N)+
\lambda_{N^-}^2 \frac{\gamma_0 + 1}{2} \delta(q_0-M_{N^-}) + 
{\rm continuum}\ .
\end{eqnarray}
Then we derive the following sum rule from Eq.(\ref{jkos}) with $W(q_0)=1$
\begin{eqnarray}
A(s_+)-B(s_+)&=&\lambda_N^2\label{over1}\ , \\
A(s_-)+B(s_-)&=&\lambda_{N^-}^2 \label{over2}\ ,
\end{eqnarray}
where in our case,
\begin{eqnarray}
A(s_{\pm})&=&\int_0^{\sqrt{s_{\pm}}} dq_0 \Big[{ q_0^7  \over 2^{10}
  \times 3^2 
                    \times  5 \times \pi^3 } (21 t^2 +10 t + 21)\nonumber \\
&&+{q_0^3  \over 2^{11}\times 9\times \pi } (45 t^2 + 10t + 45)  
                 \langle \frac{\alpha_s}{\pi} {\cal G}^2 \rangle  
                   +{1 \over 12 } \pi \delta (q_0) \langle {\bar q} q \rangle
                        \langle g_s {\bar q} \sigma \cdot {\cal G} q \rangle
                        (t^2 +t) \Big]  \ ,\\
B(s_{\pm})&=&\int_0^{\sqrt{s_{\pm}}} dq_0  \Big[- {q_0^4 \over 2^5 \pi} 
              (t^2-1)
                    \langle {\bar q} q \rangle + { q_0^2  
                    \over 2^7 \pi} 
       (3 t^2 + 2 t -5) \langle g_s {\bar q} \sigma \cdot {\cal G} q \rangle\ 
\nonumber \\
          &&-{\pi (t^2-1) \over 192} \langle \frac{\alpha_s}{\pi} 
           {\cal G}^2 \rangle \langle {\bar q} q \rangle \Big]\ .
\end{eqnarray}
Here we fix the positive-parity continuum threshold to  
the second lowest resonance of its kind,  $s_+ = (1.44\ {\rm GeV})^2$, and 
for the negative-parity threshold, we take $s_- = (1.65\ {\rm GeV})^2$ as it 
is the next higher resonance with the negative-parity.

Eq.(\ref{over1}) is a kind of duality relation between the quark and nucleon
state.  It equates the integrated positive-parity spectral strength within 
the duality interval ($0 \sim \sqrt{s_+}$)  of the OPE part to that of the
hadronic counterpart, which is the coupling strength of the nucleon.
Note that the right hand sides of Eqs.~(\ref{over1}), (\ref{over2})
are positive definite.
Now, we look for the optimal $t$ which makes the left hand side
of Eq.~(\ref{over1}) zero while maximizing 
$\lambda_{N^-}^2$ from Eq.~(\ref{over2}).   Our expectation is that
such $t$ yields the
interpolating field which couples strongly to the negative-parity baryon, 
specifically S$_{11}$.
Since the quark-gluon condensate is not well known, we determine $t$ for
the following quark-gluon parameter,  
$$\lambda^2_q =
 0.4,\ 0.5, \ 0.6, \ 0.7,\ 0.8,\ 0.9,\ 1\  {\rm GeV}^2\ .$$  
These values are arbitrary chosen from
the acceptable range of $\lambda^2_q$
between $0.4 -1$ GeV$^2$.  [See Eq.~(\ref{qgc}) for 
the definition of
$\lambda^2_q$.]  With the following parameter set, 
\begin{eqnarray}
&&\langle {\bar q} q \rangle = -(0.23\ {\rm GeV})^3\, \\
&&\langle {\alpha_s \over \pi} {\cal G}^2 \rangle = (0.35\ {\rm GeV})^4\ ,
\end{eqnarray}
we obtain the optimal $t$ for each $\lambda^2_q$ which are 
listed in Table~\ref{tab}. Note that our result is not sensitive to 
the quark or gluon condensate within the
error bars as indicated in Eqs.~(\ref{qq}) and (\ref{gluon}). 
In general, there are two values of $t$ which make the left-hand side of 
Eq.~(\ref{over1}) zero.  We choose $t$ which yields larger value of 
$\lambda^2_{N^-}$ in Eq.~(\ref{over2}).

With this optimal $t$ value, we are now in a position to 
calculate $M_{N^-}$.    Substituting all the parameters into Eq.~(\ref{mass})
we obtain the Borel curve for $M_{N^-}$. Some of them are shown 
in Fig.~\ref{Borel}.  The solid line is for $\lambda^2_q=0.4$ GeV$^2$,  
the dashed line is for $\lambda^2_q=0.6$ GeV$^2$, and the dot-dashed line is
for $\lambda^2_q=1$ GeV$^2$ . Note that all three curves 
have the stable region from which the spectral mass can be obtained. 
At $\lambda^2_q=0.6$ GeV$^2$, our prediction for
$M_{N^-}$ obtained from the stable Borel plateau is 1.59 GeV.  Similarly,
at $\lambda^2_q=1$ GeV$^2$, our prediction is  $M_{N^-}=1.42$ GeV. 
All other predictions for $M_{N^-}$ with different $\lambda_q^2$ values 
lie within these two limits as shown in Table~\ref{tab}. 
The best prediction for S$_{11}$ mass is obtained for 
$\lambda^2_q=0.435$ GeV$^2$ or $\lambda^2_q=0.69$ GeV$^2$.  However,
for $\lambda^2_q = 1 $ GeV$^2$, our prediction is off by 7\%
from the experimentally known S$_{11}$ mass, 1.535 GeV.   To make more
reliable prediction, it is important to narrow down the range  
of the quark-gluon condensate value from other studies.  However, it should 
be emphasized that our prediction is obscured only by 7\% from the
barely known quark-gluon condensate.   
That is, for a given value of $\lambda^2_q$, the optimal
procedure that we have employed chooses the
interpolating field which couples very strongly to S$_{11}$.

As we have succeeded in predicting the S$_{11}$ at least in qualitative level,
it would be interesting to understand the origin of the large
mass splitting between S$_{11}$ and nucleon.  It should be noted that 
the dominant portion of our prediction comes from the dimension-five 
quark-gluon operator.  The contribution from this dimension-five operator is 
enhanced by the covariant derivative
acted on the d-quark of the nucleon interpolating field because the covariant
derivative modifies the quark configuration to increase the average virtual
momentum of the current quark. 
However, this is not the case for the nucleon sum rule 
where the quark configurations are mostly in the s-wave state. 
Indeed, with {\it Ioffe}'s choice for the nucleon interpolating field,  
the quark-gluon condensate does not enter in the prediction of
nucleon mass.  Therefore, within our picture, the large mass splitting 
between nucleon and S$_{11}$ is mainly driven by the enhancement of the 
quark-gluon 
condensate  which increases the contributions from 
the chiral-symmetry breaking terms.

Another interesting feature of our result is that, for our current, we can
always find suitable linear combination
of currents such that it has a strong coupling with the negative-parity baryon
and  zero coupling to the positive-parity baryon.  This justifies the use of
Eq.(\ref{mass}) or any other sum rule constructed with this current under the
assumption that the lowest mass resonance that couple to this current is the
S$_{11}$.
This is not always possible for any other interpolating field for the nucleon,
especially when the current contains no derivative.  For example, using
Eq.(\ref{nucl1}) as an interpolating field, it is not possible to find any
value of $t$ such that the current has a strong coupling
to the negative-parity baryon and zero coupling to the  positive-parity
baryon.  Therefore, our interpolating field
can be used to construct any conventional sum rule to study
other properties of the S$_{11}$ relevant
in reactions involving electromagnetic or hadronic probes\cite{hung}.

In summary, we have introduced a new interpolating field for S$_{11}$ 
motivated by the bag model or nonrelativistic quark model.  Using the 
interpolating field, we have calculated the correlation function in the 
operator product expansion up to dimension eight.  To predict S$_{11}$ mass,
we have  optimized the linear combination at a given QCD parameter
$\lambda^2_q$ while fixing other well known parameters.   A good 
prediction for S$_{11}$ mass is obtained under this formalism and
its value contains at most 7\% error due to the poorly known QCD parameter
$\lambda^2_q$.  The best value for S$_{11}$ mass is obtained for 
$\lambda^2_q =0.435\ {\rm or}\ 0.69$ GeV$^2$.

\acknowledgments
SHL was supported in part by the by the Basic Science Research
Institute program of the Korean Ministry of Education through grant no.
BSRI-95-2425, by KOSEF through the CTP at Seoul National University
and by Yonsei University Research Grant.  SHL also thanks DAAD and KOSEF for
financial support while visiting GSI.
HK  was supported by KOSEF.

\begin{table}
\caption{Our prediction for S$_{11}$ at given $\lambda^2_q$ with optimally
chosen $t$.  The listed $M_{N^-}$ is obtained from the stable Borel plateau.}

\begin{center}
\begin{tabular}{ccc}
 $\lambda^2_q$ (GeV$^2$) & Optimal $t$ & $M_{N^-}$ (GeV)  \\ 
\hline\hline
0.4 & 4.97 & 1.49 \\
0.5 & -47.52 & 1.59 \\
0.6 & -4.59 & 1.59 \\
0.7 & -2.91 & 1.53 \\
0.8 & -2.40 & 1.48 \\
0.9 & -2.21 & 1.44 \\
1.0 & -2.04 & 1.42 \\
\end{tabular}
\end{center}
\label{tab}

\end{table}

\begin{figure}
\caption{Our prediction for S$_{11}$ mass versus Borel mass.
The solid curve is for $\lambda^2_q =0.4~{\rm GeV}^2$, $t=4.97$ and the 
dashed curve is for $\lambda^2_q =0.6~{\rm GeV}^2, t=-4.59$.  
The dot-dashed curve is for $\lambda^2_q =1~{\rm GeV}^2$, $t=-2.04$.}
\label{Borel}
\end{figure}
%%%%%%%%%%%%%%%%%%%%%%%%%%%%%%%%%%%%%%%%%%%%%%%%%%%%%%%%%%%%%%%%%%%%%%%%%%

\begin{references}
\bibitem{mami}   {B. Krushce {\it et al.}, 
                            Phys. Rev. Lett. {\bf 74} (1995) 3736.}

\bibitem{elsa}   {B. Schoch {\it et al.},
                            Acta Phys. Polonica {\bf 24} (1993) 1765.}

\bibitem{eta_the}   {M. Benmerrouche and Nimai C. Mukhopadhyay,
                            Phys. Rev. Lett. {\bf 67} (1991) 1070;
                     Zhenping Li,
                            Phys. Rev. D {\bf 52} (1995) 4961;
                     N. Kaiser, P. B. Siegel, and W. Weiss,
                            Phys. Lett. B {\bf 362} (1995) 23;
                     Ch. Sauermann, B. L. Friman, and W. N\"orenberg,
                            Phys. Lett. B {\bf 341} (1995) 261;
                     M. Benmerrouche, Nimai C. Mukhopadhyay, and J. F. Zhang,
                            Phys. Rev. D {\bf 51} (1995) 3237.}
\bibitem{SVZ}     {M.A. Shifman, A.I. Vainshtein, and V.I. Zakharov,
                            Nucl. Phys. {\bf 147} (1979) 385,448.}
\bibitem{qsr}     {For general review see L.J. Reinders, H. Rubinstein and 
                    S. Yazaki, Phys. Rep. {\bf 127} (1985) 1.}
\bibitem{Tom1}     {T.D. Cohen, R.J. Furnstahl, D.K. Griegel and X. Jin, 
                    Prog. Part. Nucl. Phys. {\bf 35} (1995) 221.}
\bibitem{Lee1}     {T. Hatsuda and Su H. Lee, Phys. Rev. {\bf C46}(1992) R34,
                    T. Hatsuda, Su H. Lee and H. Shiomi, Phys. Rev. {\bf C52}
                   (1995) 3364.}
\bibitem{jido}      {D. Jido, N. Kodama, and M. Oka,
                     Los Alamos Preprint hep-ph/9604280.}
\bibitem{hung}     {Hungchong Kim and S.H. Lee in preparation.}
\bibitem{ioffe}     {B. L. Ioffe,
                            Nucl. Phys.  {\bf B188} (1981) 317.}
\bibitem{Esp}     {D. Espriu, P. Pascual and R. Tarrach,
		              Nucl. Phys. {\bf B214} (1983) 285.}
\bibitem{griegel}   {David K. Griegel,
                     Ph.D thesis,
                     University of Maryland (1991).}
\bibitem{chodos}    {A. Chodos, R. L. Jaffe, K. Johnson, and C. B. Thorn,
                            Phys. Rev. D {\bf 10} (1974) 2599.}
\bibitem{isgur}     {Nathan Isgur and Gabriel Karl,
                            Phys. Lett. B {\bf 72} (1977) 109. }
\bibitem{Ovc}       {A. A. Ovchinnikov and A. A. Pivovarov, 
                            Yad. Fiz. {\bf 48} (1998) 1135.}
\bibitem{bak}       {A. P. Bakulev and A. V. Radyushkin,
                            Phys. Lett. B {\bf 271} (1991) 223. }
\bibitem{kremer}    {M. Kremer and G. Schierholz,
                            Phys. Lett. B. {\bf 194} (1987) 283.}
        

\end{references}
\end{document}